\documentclass[aps,prl,reprint,groupedaddress]{revtex4-2}
\usepackage[normalem]{ulem}
\usepackage{amsmath}
\usepackage[dvipsnames]{xcolor}
\usepackage{amssymb}
\usepackage{graphicx}
\usepackage{txfonts}
\usepackage{cancel}
\usepackage{tikz-feynman} 
\usepackage{hyperref}
\hypersetup{
  colorlinks = true, 
  urlcolor   = blue, 
  linkcolor  = blue, 
  citecolor  = blue  
}
\usepackage{enumitem} 
\definecolor{blue}{RGB}{0,0,220}


\begin{document}

\title{Physics of the QCD Trace Anomaly and Renormalization in a Reducible Basis}


\author{Chen Yang}
\email{cyang127@umd.edu}

\affiliation{Department of Physics, University of Maryland, College Park, MD 20742, U.S.A.}
\date{\today}

\begin{abstract}

The QCD trace anomaly is a well-established textbook result in quantum field theory with several prominent features: (1) it arises from the quantum breaking of scale symmetry at ultraviolet (UV) scales, yet its full renormalized expression is independent of the particular UV regulator used, whether lattice or dimensional regularization; (2) although nominally of order $\alpha_s$, it is free of renormalization-scheme ambiguity; and (3) it is free of UV divergences and therefore scale independent. We show that renormalization in a reducible operator basis introduces additional convention dependence in its separation into quark and gluon parts, which do not share the properties of the whole anomaly itself. An irreducible decomposition instead keeps the trace anomaly as a single, convention-free component, and gives transparent physical meanings to anomaly-related nucleon physics, including the nucleon mass structure and internal forces.


\end{abstract}

\maketitle

\section{Introduction}

The physics of trace anomaly has been understood through development of renormalized quantum field theory in the 1970s, and has now become a well-known textbook material~\cite{Peskin:1995ev,Schwartz:2014sze,Collins:1984xc}. Early studies showed that the dilatation Ward identities of a classically scale-invariant theory can acquire anomalous contributions~\cite{Callan:1970ze,Coleman:1970je}. This is because the divergent quantum fluctuations break the classical scale symmetry by a new scale introduced through the UV regularization, such as $\mu$ in the dimensional regularization~\cite{tHooft:1972tcz,Bollini:1972ui} or lattice spacing $a$~\cite{Rothe:1992nt}. The associated Noether's current, the dilatation current $J_D^{\mu}$~\cite{Noether1918,Blaschke:2016ohs}, acquires a non-zero divergence, $\partial_\mu J^{\mu}_D=T^{\mu}_{\ \mu}\neq0$, thus inducing an additional trace term to the energy-momentum tensor (EMT), i.e., the so-called trace anomaly. In the Abelian and non-Abelian gauge theories, the EMT is formally written as~\cite{rosenfeld1940energy,BELINFANTE1940449}, 
\begin{equation}
    {T}^{\mu\nu} = \bar{\psi}i\overleftrightarrow{\mathcal{D}}^{(\mu}\gamma^{\nu)}\psi+ \frac{g^{\mu\nu}}{4}F^{2}-F^{\mu\alpha}F_{\ \alpha}^{\nu} \ ,
\end{equation}
where $\overleftrightarrow{\cal D} = (\overrightarrow{\cal D}-\overleftarrow{\cal D})/2$, $a^{(\mu}b^{\nu)} = (a^\mu b^\nu + a^\nu b^\mu)/2$ and a sum over fermion flavors is implied. Studying its renormalization leads to the standard operator expression for the trace anomaly in quantum chromodynamics (QCD)~\cite{Joglekar:1975nu,Adler:1976zt,Collins:1976yq,Nielsen:1977sy}, 
\begin{equation}
    T^{\mu}_{\ \mu} = (1+\gamma_m) m\bar\psi\psi + \frac{\beta\left(g\right)}{2g}(F^{2})_R \equiv m\bar\psi\psi+ T^{\mu}_{a\mu} \label{eq:Anomaly}
\end{equation}
where $m$ is the quark mass, $g$ is the strong-coupling constant, $\gamma_m=-\frac{{\rm d}\log m}{{\rm d}\log \mu}$ is the mass anomalous dimension, $\beta=\frac{{\rm d}g}{{\rm d}\log \mu}$ is the beta function and $\mu$ is the renormalization scale. As a fundamental result, the trace anomaly is important in understanding the non-perturbative dynamics in QCD.


When studying the impact of the trace anomaly on the nucleon structures, its several important characteristics must be respected. Firstly, at the UV scale, the trace anomaly takes different forms as (evanescent) bare operators under different UV regulators such as dimensional, Pauli-Villars or lattice regularizations~\cite{Collins:1984xc}. Despite this, different regulators yield the same anomalous Ward-Takahashi identity (WTI) and the same finite anomaly term at the physical IR scale~\cite{Ji:1994av,Ji:1995sv,Caracciolo:1989pt,Caracciolo:1991cp}. 
A similar phenomenon happens for the axial anomaly~\cite{Jackiw:1968xt,PhysRev.182.1517} for which the same result 
has been derived under: 1) dimensional regularization in the triangle diagrams~\cite{Collins:1984xc,Peskin:1995ev,Schwartz:2014sze}; 2) a small spacetime separation in the axial current and its divergence~\cite{Schwinger:1962tp}; 3) a UV-suppressed exponential factor in the anomalous WTI~\cite{Fujikawa:1979ay,Fujikawa:1983bg}; 4) lattice spacing $a$ in the anomalous WTI~\cite{Wilson:1974sk,Karsten:1980wd,Kawai:1980ja,Sharatchandra:1981si,Ginsparg:1981bj,Kerler:1981tb,Seiler:1981jf,Fujikawa:1983az,Rothe:1992nt}. Therefore, the physical meaning of the trace anomaly cannot be (uniquely) traced back to the details of what happened at the UV scale. Instead, it is manifested through 
the anomaly operator itself at the IR scale. 

The anomaly nominally starts with the order of $\alpha_s$ coupling. In standard renormalization of composite operators, order-of-$\alpha_s$ terms are subtraction scheme dependent. However, this is not the case with the anomaly. At UV scale, the anomaly operator is an evanescent operator and vanishes in the classical limit~\cite{Collins:1984xc}, and thus there is no order-of-1 contribution as in Eq.~(\ref{eq:Anomaly}). The leading $\alpha_s$ contributions in $\beta(g)$ and $\gamma_m$ are clearly scheme independent. This can easily be overlooked if they are renormalized together with operators with tree level contributions where the ambiguities from the finite-term subtractions also start at order ${\cal O}(\alpha_s)$. Therefore, the trace anomaly shall be treated separately from the renormalization of the regular operators, i.e., the traceless parts of the QCD EMT. 

Finally, the whole trace anomaly is renormalization-group (RG) invariant, which can be understood through the QCD EMT, a conserved current free of UV divergences and overall renormalization. Its traceless and trace parts, being two irreducible representations of the Lorentz symmetry group, are thus also UV finite. The trace part contains both a quark mass and trace anomaly contribution as in Eq.~(\ref{eq:Anomaly}). Since the quark mass contribution $m\bar{\psi}\psi$ is RG invariant, the trace anomaly is also free from scale evolutions. 

An important consequence of the above discussions is that the QCD EMT cannot be simply separated into a sum of quark and gluon contributions through renormalizing the bare operators in $T^{\mu\nu}=T_{q,B}^{\mu\nu}+T_{g,B}^{\mu\nu}$,
because the trace anomaly itself cannot be traced back uniquely to separate quark and gluon UV operators. The resulting renormalized quark and gluon EMTs $T^{\mu\nu}_{q/g,R}$ thus have no unique physical meanings. The natural physical organization is to identify the anomalous contributions through their renormalized forms, namely $\gamma_m m\bar{\psi}\psi$ and $[\beta(g)/(2g)]F^2$ as the quark and gluon components, respectively, as was done in Ref. \cite{Ji:1994av}. 

This distinction concerns the recent reducible-basis renormalization of the QCD EMT introduced in Refs.~\cite{Makino:2014taa,Harlander:2018zpi,Hatta:2018sqd,Tanaka:2018nae} and subsequent interpretations of the nucleon mass sum rule~\cite{Lorce:2017xzd,Lorce:2018uyy,Ahmed:2022adh,Rodini:2020pis,Metz:2020vxd,Lorce:2021xku,Tanaka:2025hhv,Bollweg:2026nnr}. In this paper, we study the {\it reducible operator basis} renormalization and examine if it fully considers the above features of the trace anomaly. For simplicity, we consider the chiral limit in the rest of the paper. 


\section{Mixing Trace Anomaly and Regular Operators in Reducible Operator Basis} 




Renormalization of composite operators involves choosing an {\it operator basis}, with each operator ${\cal O}_{i,B}$ in the basis renormalized subject to mixing, manifested as the renormalization-factor matrix $\{Z_{ij}\}$,
\begin{equation}
    O_{i,R} = \lim_{\epsilon\to 0}\sum_{j}Z_{ij} O_{j.B} \ .
\end{equation}
An important strategy is to isolate operators with different UV structures and physical symmetries. This is achieved by using the irreducible representations of the symmetry groups as the operator basis, generally acknowledged as the {\it standard renormalization procedure} or a physics-motivated renormalization scheme, where $\{Z_{ij}\}$ becomes block-diagonal and each block represents different quantum numbers~\cite{Collins:1984xc}. 

For the QCD EMT, $\{Z_{ij}\}$ contains two blocks: the traceless ($\bar{T}^{\mu\nu}$) and trace ($\hat{T}^{\mu\nu}$) parts, 
which belong to two different irreducible representations of Lorentz group, 
\begin{align}
    \bar{T}^{\mu\nu} &= T^{\mu\nu} - \frac{1}{4}g^{\mu\nu}T^{\alpha}_{\ \alpha} \equiv \bar{T}^{\mu\nu}_q + \bar{T}^{\mu\nu}_g \\
    \hat{T}^{\mu\nu} &= \frac{1}{4}g^{\mu\nu}T^{\alpha}_{\ \alpha}
\end{align}
The trace part is the purely anomaly operator in the massless case and arises from quantum physics, while the traceless part contains classical quark and gluon contributions that undergo mixing under the renormalization, 
\begin{equation}
    \renewcommand{\arraystretch}{1.3}
    \begin{pmatrix}
        \bar{T}_{q,R}^{\mu\nu}(\mu) \\
        \bar{T}_{g,R}^{\mu\nu}(\mu) \\
        \hat{T}^{\mu\nu}_R
    \end{pmatrix}
    =
    \begin{pmatrix}
        Z_{qq} & Z_{qg} & 0 \\
        Z_{gq} & Z_{gg} & 0 \\
        0      & 0      & 1
    \end{pmatrix}
    \begin{pmatrix}
    \bar{T}_{q}^{\mu\nu} \\
    \bar{T}_{g}^{\mu\nu} \\
    \hat{T}^{\mu\nu}
    \end{pmatrix} 
    \label{eq:Ji-EMT}
\end{equation}
where $\mu$ is the renormalization scale. These renormalization factors have been perturbatively calculated to four-loop order~\cite{Moch:2021qrk}. Lorentz symmetry gives no reason to mix the renormalization of the two parts.

In the unconventional scheme introduced in Refs.~\cite{Makino:2014taa,Harlander:2018zpi}, a reducible basis of the Lorentz group is adopted, in which operators of different Lorentz structures are mixed. Besides, the trace anomaly is not separated from regular operators but is renormalized together, thus introducing additional ambiguities. To explain this, we consider the following more general reducible basis that allows us to unify different schemes in Refs.~\cite{Makino:2014taa,Harlander:2018zpi,Lorce:2017xzd,Lorce:2018uyy,Hatta:2018sqd,Tanaka:2018nae,Tanaka:2025hhv,Ahmed:2022adh,Rodini:2020pis,Metz:2020vxd,Lorce:2021xku} in dimensional regularization $d=4-2\epsilon$,
\begin{equation}
    \renewcommand{\arraystretch}{1.5}
    \begin{pmatrix}
        -F^{\mu\alpha}F^{\nu}_{\ \alpha}\\
        c_2g^{\mu\nu} F^2\\
        \bar{\psi}i\overleftrightarrow{\mathcal{D}}^{(\mu}\gamma^{\nu)}\psi
    \end{pmatrix}_R
    = \lim_{\epsilon\to0}
    \begin{pmatrix}
        Z_{T} & Z_{M} & Z_{L} \\
        0 & Z_{F} & 0\\
        Z_{Q} & Z_{B} & Z_{\psi}
    \end{pmatrix}
    \begin{pmatrix}
        -F^{\mu\alpha}F^{\nu}_{\ \alpha}\\
        c_2g^{\mu\nu} F^2\\
        \bar{\psi}i\overleftrightarrow{\mathcal{D}}^{(\mu}\gamma^{\nu)}\psi
    \end{pmatrix}
    \label{eq:Hatta-R}
\end{equation}
where $c_2 = 1+c_0(1-d/4)$ with $c_0$ a constant of order 1, and $Z'$s are defined through minimal subtraction. By perturbative calculations or the same procedure as Ref.~\cite{Hatta:2018sqd}, the resulting renormalized quark and gluon EMTs are given by a mixture of traceless and trace anomaly operators, 
\begin{align}
    T^{\mu\nu}_{q,R}(\mu) & = \bar{T}^{\mu\nu}_{q,R}(\mu) + \frac{1}{4}g^{\mu\nu}x(F^2)_{R} \ , \label{eq:rrqEMT}\\
    T^{\mu\nu}_{g,R}(\mu) & = \bar{T}^{\mu\nu}_{g,R}(\mu) + \frac{1}{4}g^{\mu\nu}\left(\frac{\beta(g)}{2g}-x\right)(F^2)_{R} \ . \label{eq:rrgEMT}
\end{align}
This leaves the total trace anomaly intact but introduces additional conventions in its separation. These traces have been interpreted as its ``quark and gluon'' contributions, respectively. Different choices of the reducible basis through $c_0$ recover the undetermined scheme dependence $x$ proposed by Metz, Pasquini, Rodini and collaborators (MPR)~\cite{Rodini:2020pis,Metz:2020vxd,Lorce:2021xku}. At one-loop order,
\begin{equation}
    x = \frac{\alpha_s}{4\pi} \frac{N_f}{3} (1-c_0) + {\cal O}(\alpha_s^2)\ , \label{eq:x-c0}
\end{equation}
where $N_f$ is the quark flavor number. See Appendix~\ref{appsec:renorm} for detailed derivations. Choosing $c_0=1$ and $c_0=0$ reduce to the schemes of Lorc\'e~\cite{Lorce:2017xzd,Lorce:2018uyy} and Hatta, Rajan, Tanaka and their collaborators (HRT)~\cite{Hatta:2018sqd}, respectively. At one loop, all these different schemes are thus unified through the choice of the reducible basis $c_0$. 

Although the perturbative results of Refs.~\cite{Makino:2014taa,Harlander:2018zpi,Hatta:2018sqd,Tanaka:2018nae} are technically correct and, for the total EMT, equivalent to Eq.~(\ref{eq:Ji-EMT}), the physical interpretation of the separated traces remains unclear. Such separation has introduced an additional convention into the otherwise scheme-independent quantity. As we discuss below, this new scheme-dependence is neither necessary nor physically straightforward. 

Firstly, such renormalization breaks the irreducible Lorentz structure. The bare $d$-dimensionally traceless quark operator has traceless counterterms within the spin-two sector, whereas the renormalized tensor in Eq.~(\ref{eq:rrqEMT}) contains an additional finite scalar term. Clearly, this term is an artifact of the arbitrariness (choice of the coefficient in the front of $g^{\mu\nu}$ when adopting minimal subtraction) of the reducible basis. The irreducible Lorentz structure is preserved for $c_2=4/d$ which gives $x=0$. At one loop, this corresponds to $c_0=1$. See Appendix~\ref{appsec:renorm} for details. In other words, one associates $g^{\mu\nu}$ with $1/d$ factor in dimensional regularization and minimal subtraction because the trace produces a factor of $d$, much as a gluon has $d-2$ physical polarizations in dimensional regularization~\cite{Collins:1984xc}. While this particular choice does lead to the standard result that the resulting renormalized gluon EMT contains the entire anomaly, such attribution shall not be made through a renormalization scheme, as explained in Introduction. It is a physical statement that the anomaly is reflected through the IR gluon physics. The scalar gluon from the trace anomaly has distinct physical origins from the traceless-tensor gluon and simply viewing them as an overall gluon contribution obscures this distinction. 


The complication grows with tensor rank. For higher-rank tensors, the reducible basis leads to more significant mixing effects and a more complicated renormalization matrix. For example, the parity-even dimension-six pure-gluon sector relevant to symmetric rank-four tensors, modulo total derivatives and retaining equation-of-motion operators, contains one traceless rank-four, three traceless rank-two, and two scalar multiplets. In an irreducible basis, Lorentz symmetry organizes their renormalization into $1\times1$, $3\times3$, and $2\times2$ blocks, with at most $14$ coefficients. In a reducible basis analogous to Eq.~(\ref{eq:Hatta-R}), the mixing matrix is only block-triangular, with up to 25 coefficients. See Appendix~\ref{appsec:rank4} for details. An irreducible basis makes these constraints manifest, allows the three sectors to be treated separately and significantly reduces the complication of mixing effects. 

On the lattice, the resulting mixing effects expose the calculation to power divergences ($a^{-n}$) from lower-dimension operators that would otherwise be symmetry-forbidden~\cite{Gockeler:1996mu}. In the irreducible basis, the traceless EMT operators cannot mix with lower-dimensional operators. The trace channel mixes with the identity operator, giving an $a^{-4}$ vacuum contribution. The reducible-basis quark and gluon EMTs contain trace components and therefore inherit this $a^{-4}$ mixing. Power-divergent mixing requires a careful nonperturbative subtraction, as is familiar from the linear divergence of Wilson-line operators~\cite{Ishikawa:2016znu,LatticePartonLPC:2021gpi}. The gradient flow method handles them by defining finite operators at positive flow time~\cite{Makino:2014taa,Harlander:2018zpi,Bollweg:2026nnr}. More seriously, when operators with different quantum numbers are mixed, they lose connection to experimental measurements. In the present case, while the traceless quark and gluon EMTs are related to measurable observables in deep-inelastic scatterings, the new renormalized quark and gluon operators in Eqs.~(\ref{eq:rrqEMT}) and (\ref{eq:rrgEMT}) involve additional convention $x$, which is not fixed by experiments. 

\section{Trace Anomaly in Nucleons}

Based on these observations of the trace anomaly as well as the Lorentz symmetry group, the QCD EMT thus has two components in the chiral limit: the traceless component with classical tree-level contributions dressed by quantum fluctuations and the trace anomaly as a purely quantum component. By further decomposing the traceless part into quarks and gluons, the three EMT operators in Eq.~(\ref{eq:Ji-EMT}) are explicitly,
\begin{align}
    \bar{T}^{\mu\nu}_{q,R} & = [\bar{\psi}i\overleftrightarrow{\mathcal{D}}^{(\mu}\gamma^{\nu)}\psi]_R \ , \\
    \bar{T}^{\mu\nu}_{g,R} & = \left[\frac{g^{\mu\nu}}{d}F^{2}-F^{\mu\alpha}F_{\ \alpha}^{\nu}\right]_R\ , \\
    \hat{T}^{\mu\nu}_a & = \frac{g^{\mu\nu}}{4}\frac{\beta\left(g\right)}{2g}(F^{2})_R  \ ,
\end{align}
Further information on the nucleon structures can be extracted based on this decomposition. In this section, we briefly review the nucleon mass sum rule~\cite{Ji:1994av,Ji:1995sv,Ji:2021pys,Ji:2021mtz,Ji:2021qgo,Rothe:1995av} and the color-Lorentz force on quarks within the proton~\cite{Ji:2025gsq,Ji:2025qax,Ji:2026lyj,Yu:2026uuf}. 

\subsection{Nucleon Mass Sum Rule}

According to Einstein's mass–energy equivalence principle~\cite{einstein1905does}, the nucleon mass is defined as its rest-frame energy, i.e., the expectation values of the Hamiltonian operator in the nucleon zero-momentum eigen-state, 
\begin{equation}
    M \equiv \frac{\langle P | \int {\rm d}^3 \vec{x}~T^{00}(x)|P\rangle}{\langle P|P \rangle} \bigg{|}_{\vec{P}=0} \equiv \langle T^{00} \rangle_{\vec{P}=0} \ ,
\end{equation}
where $M$ is the nucleon mass, $P^\mu$ is the four momentum of the nucleon state normalized as $\langle P^{\prime}|P \rangle=2P^0(2\pi)^3\delta^{(3)}(\vec{P}^{\prime}-\vec{P})$ and the subscript represents the momentum of the external state. Based on the above decomposition of the QCD EMT, the total Hamiltonian density has the following three components, 
\begin{align}
    \bar{\cal H}_q &\equiv \bar{T}^{00}_{q,R} \equiv (\bar{\psi}i\vec{\cal D}\cdot\vec{\gamma}\psi)_R  \ , \label{eq:Ebarq} \\
    \bar{\cal H}_g &\equiv \bar{T}^{00}_{g,R} \equiv  \frac{1}{2}(\vec{E}^{2}+\vec{B}^{2})_R \ , \label{eq:Ebarg} \\
    \hat{\cal H}_a &\equiv \hat{T}^{00}_a \equiv -\frac{\beta(g)}{4g}(\vec{E}^2-\vec{B}^2)_R \ , \label{eq:Ehata}
\end{align}
where $\{\vec{E},\vec{B}\}$ are the chromo-electromagnetic fields and the subscript $R$ represents renormalized operators. The expectation values of these three Hamiltonian density components in the rest-frame nucleon are, 
\begin{align}
    \hat{M}_a &= \langle{\cal H}_a\rangle_{\vec{P}=0} = \frac{1}{4} M \ , \\
    \bar{M}_q &= \langle{\cal H}_q\rangle_{\vec{P}=0} = 3\langle x\rangle_q(\mu)\hat{M}_a\ ,\\
    \bar{M}_g &= \langle{\cal H}_g\rangle_{\vec{P}=0} = 3\langle x\rangle_g(\mu)\hat{M}_a \ ,
\end{align}
where $\langle x\rangle_{q,g}(\mu)$ are the second moments of the quark and gluon PDFs, representing the momentum fractions carried by quarks and gluons. These fractions can be measured through experiments such as the deep-inelastic scattering~\cite{Hou:2019efy} as well as evaluated through lattice QCD calculations~\cite{Abdel-Rehim:2016won,Alexandrou:2017oeh,Yang:2018nqn,Alexandrou:2020sml,Hackett:2023rif,Bollweg:2026nnr}. 

Note that the traceless parts contribute three times the trace anomaly to the total nucleon mass $\bar{M} = 3\hat{M}$. This fixed ratio follows from the Lorentz structure of a one-particle forward matrix element in four dimensions, which has been called the Virial theorem~\cite{Ji:1994av,Ji:1995sv}. Remarkably, this indicates that the trace anomaly serves as an emergent IR scale for generating the QCD nucleon mass. The same phenomenon has also been observed in the lattice gauge theory~\cite{Rothe:1995av}. 

These three components admit transparent physical meanings as different energies. $\bar{\cal H}_q$ and $\bar{\cal H}_g$ are the kinetic and potential energies of quarks and gluons. Both have classical tree-level counterparts, dressed by quantum fluctuations, which can thus be understood as the ``classical energy''. $\hat{\cal H}_a$ is a quantum anomalous energy (QAE) representing a purely quantum effect~\cite{Ji:2021pys,Ji:2021mtz,Ji:2021qgo}, arising from the scalar gluon field. Heuristically, it resembles the bag constant of bag-model phenomenology and the Casimir effect. 

These interpretations are supported by the Ward-Takahashi identity of the Poincar\'e symmetry~\cite{Rothe:1995av,Ji:2021pys,Ji:2021mtz,Ji:2021qgo}. By rescaling the time direction, which is treated as a time translation and thus yields the full Hamiltonian, 
\begin{equation}
    H = H_c + H_a = \int {\rm d}^{3}\vec{x}~\bar{T}^{00}(\vec{x}) + \frac{1}{4} g^{00}\int {\rm d}^{3}\vec{x}~T^{\mu}_{\ \mu}(\vec{x}) \ ,\label{eq:H}
\end{equation}
The first term $H_c$ is the normal ``classical'' energy and the second term $H_a$ is the emergent QAE. These exactly coincide with traceless part and the trace anomaly. Therefore, the trace anomaly does contribute to the Hamiltonian, in contrast to the interpretation in Refs.~\cite{Rodini:2020pis,Metz:2020vxd,Lorce:2021xku} that it appears only in the spatial components of the EMT.

We also note alternative methods to study the nucleon mass structure through the trace of the QCD EMT, or the invariant mass relation $P^\mu P_\mu=M^2$~\cite{Hatta:2018sqd,Liu:2021gco}. However, the trace provides an identity for computing the nucleon mass rather than an energy decomposition. Although its matrix element splits linearly into operator contributions, these contributions are not energies, and invariant mass itself is not additive for a composite system. The trace decomposition therefore does not give a mass sum rule whose terms carry the physical meaning of the whole.

\subsection{Confinement Force}

The QCD EMT encodes the energy, momentum and the momentum-current densities carried by different components. Analogous to Newton's law, the divergence of the quark EMT, or classically the momentum change of a moving particle, yields the force components. 

To illustrate the physical meaning of this divergence, consider the classical limit and use the equations of motion, 
\begin{align}
    \partial_{\mu}\bar{T}_{q}^{\mu j}
    &=gF_{a}^{\mu j}\bar{\psi}\gamma_{\mu}t_{a}\psi=g\rho_{a}E_{a}^{j}+g\left(\vec{j}_{a}\times\vec{B}_{a}\right)^{j}\equiv {\cal F}^j_q
\end{align}
where $(\rho_a,\vec{j}_a)=\bar{\psi}\gamma^{\mu}t_{a}\psi$ is the color current. The quantum operator relation is derived in Ref.~\cite{Tanaka:2018wea}, which concerns single insertions of the EMT and its divergence between physical, on-shell nucleon states, in which the equation-of-motion and BRST-exact operators vanish; see also~\cite{Polyakov:2018exb,Won:2023zmf,Freese:2024rkr}. Therefore, this divergence gives exactly the standard {\it color-Lorentz force} density, weighted by quark probability density. This classical relation motivates the interpretation of the quark-EMT divergence as a force density. 

By using the conservation law, there are two components in this color-Lorentz force in QCD, 
\begin{equation}
    \partial_\mu \bar{T}^{\mu i}_q = - \partial_\mu \bar{T}^{\mu i}_g - \partial_\mu \hat{T}^{\mu i}_a \equiv {\cal F}_{g}^i + {\cal F}_{a}^i \ ,
\end{equation}
These force density components can be obtained by Fourier-transforming the off-forward nucleon matrix elements of the QCD EMTs in the Breit frame~\cite{Sachs:1962zzc,Polyakov:2018zvc,Jaffe:2020ebz} or the infinite-momentum frame~\cite{Dirac:1949cp,Weinberg:1966jm,Kogut:1969xa,Burkardt:2002hr,Miller:2007uy,Miller:2025zte}.  Under this irreducible decomposition, the above force densities are related to the EMT form factors $\{A_{q,g},B_{q,g},C_{q,g}\}$~\cite{Ji:2021mtz}, which can be accessed through deep exclusive processes~\cite{Ji:1996nm,Ji:1996ek,Hatta:2018ina,Hatta:2018sqd,Hatta:2019lxo,Mamo:2019mka,Guo:2021ibg,Mamo:2022eui,Burkert:2018bqq,Kumericki:2019ddg,Duran:2022xag}, lattice QCD calculations~\cite{Shanahan:2018pib,Shanahan:2018nnv,Pefkou:2021fni,Hackett:2023nkr,Hackett:2023rif} as well as theoretical and phenomenological studies~\cite{Ji:1997gm,Cebulla:2007ei,Goeke:2007fq,Goeke:2007fp,Donoghue:1991qv,Pasquini:2014vua,Broniowski:2024oyk,Cao:2024zlf,Stegeman:2025sca,Stegeman:2025tdl}. 

It has been shown in Refs.~\cite{Ji:2025gsq,Ji:2025qax,Liu:2021gco,Liu:2023cse,Liu:2026pbf,Tanaka:2025pny,Fujii:2025pkv,Fujii:2025eug} that the trace anomaly serves as a confining pressure potential and generates a large attractive force acting on quarks with its average value around $1~{\rm GeV/fm}$. It thus has a magnitude similar to the confinement string tension~\cite{Sun:2020pda,Brambilla:2021wqs}, which suggests that the trace anomaly plays a central role in quark confinement. Refs.~\cite{Ji:2026lyj,Yu:2026uuf} further report an extraction of the total force acting on quarks from experimental form-factor analyses, finding an approximately constant attractive force over intermediate distances. 

However, the reducible-basis renormalization that assigns trace anomaly to separate quark and gluon tensors introduces an additional form factor, proportional to $x$ times the form factor of $(F^2)_R$. Correspondingly, the divergence gives, 
\begin{equation}
    \partial_\mu T^{\mu\nu}_{q,R}(x) = \partial_\mu\bar T^{\mu\nu}_{q,R} + \frac{x}{4}\partial^\nu(F^2)_R
\end{equation}
Therefore, the resulting quark and gluon form factors as well as the extracted force densities depend on a convention that experiments cannot fix.

\section{Conclusion} \label{sec:conclusion}

In this paper, by highlighting the distinct features of the QCD trace anomaly, namely regulator independence, scheme independence and scale invariance, we have examined the renormalization of the QCD EMT in a reducible operator basis. Such a basis leaves the total anomaly unchanged, but the resulting separation into quark and gluon parts acquires an additional scheme dependence. This separation does not share the properties of the full anomaly and attempts to trace the UV origins of the scheme-independent IR physics through renormalization. An irreducible decomposition keeps the anomaly as a single, convention-free component, and we have reviewed how it yields transparent energy and force interpretations of nucleon structure.


\section*{Acknowledgment}
The author thanks X. Ji for suggesting this topic and useful discussions. The author also thanks Xiang Gao, Yushan Su, Kazuhiro Tanaka and Jialu Zhang for useful discussions. CY is partially supported by Maryland Center for Fundamental Physics (MCFP). 

\appendix

\section{Reducible Basis Renormalization} \label{appsec:renorm}

We use pole-only minimal subtraction with $N_f$ massless flavors and $\beta_0 = 11C_A/3-2N_f/3$. Equation-of-motion and BRST-exact operators are omitted, since they do not contribute to physical matrix elements. 

At one loop, since the change of $c_2$ in the operator basis as in Eq.~(\ref{eq:Hatta-R}) is evanescent, the divergent contributions of these bare operators are the same as those at $c_2=1$, leaving the renormalization factors unchanged as in Ref.~\cite{Hatta:2018sqd},
\begin{align}
    Z_T&=1+\frac{2N_f}{3\epsilon}\frac{\alpha_s}{4\pi},& Z_Q&=-\frac{2N_f}{3\epsilon}\frac{\alpha_s}{4\pi},\nonumber\\
    Z_\psi&=1+\frac{8C_F}{3\epsilon}\frac{\alpha_s}{4\pi},& Z_L&=-\frac{8C_F}{3\epsilon}\frac{\alpha_s}{4\pi},\nonumber\\
    Z_M&=\frac{11C_A}{12\epsilon}\frac{\alpha_s}{4\pi},& Z_B&=-\frac{N_f}{6\epsilon}\frac{\alpha_s}{4\pi},\nonumber\\
    Z_F&=1-\frac{\beta_0}{\epsilon}\frac{\alpha_s}{4\pi},& &
    \label{eq:Apoles}
\end{align}
with ${\cal O}(\alpha_s^2)$ terms suppressed. We write the bare quark and gluon EMTs as, 
\begin{align}
    T^{\mu\nu}_q &= \bar{\psi}i\overleftrightarrow{\mathcal{D}}^{(\mu}\gamma^{\nu)}\psi \\
    T^{\mu\nu}_g &= -F^{\mu\alpha}F^{\nu}_{\ \alpha}+\frac{1}{4}c_{2}g^{\mu\nu}F^{2}+c_{0}\frac{g^{\mu\nu}}{4}\frac{d-4}{4}F^{2}
\end{align}
where the last term in the gluon EMT is evanescent. Now, one may use Eq.~(\ref{eq:Hatta-R}) to express the two renormalized quark and gluon EMTs. Then, by taking the trace of the renormalized quark and gluon EMTs, one obtains,
\begin{align}
    g_{\mu\nu}T^{\mu\nu}_{q,R} &= \lim_{\epsilon\to0} \left[-Z_Q+dc_2Z_B\right]F^2 \\
    g_{\mu\nu}T^{\mu\nu}_{g,R} &= \lim_{\epsilon\to0}\left[-Z_T + dc_2\left(Z_M+\frac{Z_F}{4} \right) + \frac{c_0}{16}d(d-4)\right]F^2
\end{align}
Substituting the renormalization factors, the above equations reduce to,
\begin{align}
    g_{\mu\nu}T^{\mu\nu}_{q,R} &= \lim_{\epsilon\to0} \left[1-\left(1-\frac{\epsilon}{2}\right)\left(1+c_0\frac{\epsilon}{2}\right)\right]\frac{2N_f}{3\epsilon}\frac{\alpha_s}{4\pi}F^2 \\
    g_{\mu\nu}T^{\mu\nu}_{g,R} &= \lim_{\epsilon\to0}\left[-\frac{\epsilon}{2} - (1-c_0)\frac{N_f}{3}\frac{\alpha_s}{4\pi}\right]F^2
\end{align}
By using the relation $F^2=\left(1+\frac{\beta_0\alpha_s}{4\pi\epsilon}\right)(F^2)_R$, the two traces can be written as a finite renormalized form, 
\begin{align}
    g_{\mu\nu} T^{\mu\nu}_{q,R}(\mu) &= (1-c_0)\frac{N_f}{3}\frac{\alpha_s}{4\pi}(F^2)_R + {\cal O}(\alpha_s^2) \\
    g_{\mu\nu} T^{\mu\nu}_{g,R}(\mu) &=\left[\frac{\beta(g)}{2g}-x\right](F^2)_R + {\cal O}(\alpha_s^2)
\end{align}
By comparing these two equations with Eqs.~(\ref{eq:rrqEMT}) and (\ref{eq:rrgEMT}), one reproduces Eq.~(\ref{eq:x-c0}). 

On the other hand, we now show that the exact choice $c_2=4/d$ in Eq.~(\ref{eq:Hatta-R}), together with pole-only minimal subtraction, implies $x=0$ to all orders in perturbation theory. We begin with following basis that separates the traceless and trace sectors:
\begin{equation}
    \renewcommand{\arraystretch}{1.5}
    \begin{pmatrix}
        \bar{T}^{\mu\nu}_{g}\\ (4/d)g^{\mu\nu}F^2\\ \bar{T}^{\mu\nu}_q
    \end{pmatrix}_R
    =
    \lim_{\epsilon\to0}
    \begin{pmatrix}
        Z_{gg} & 0 & Z_{gq}\\
        0 & Z_F & 0\\
        Z_{qg} & 0 & Z_{qq}
    \end{pmatrix}
    \begin{pmatrix}
        \bar{T}^{\mu\nu}_{g}\\ (4/d)g^{\mu\nu}F^2\\ \bar{T}^{\mu\nu}_q
    \end{pmatrix}
\end{equation}
This is simply a re-arrangement of Eq.~(\ref{eq:Ji-EMT}), except that the trace channel contains an additional factor $(4/d)$. By using the exact relation $\bar{T}^{\mu\nu}_{g}=-F^{\mu\alpha}F^\nu_{\ \alpha}+(1/4)(4/d)g^{\mu\nu}F^2$, which preserves pole-only minimal subtraction, one obtains, 
\begin{equation}
    \renewcommand{\arraystretch}{1.5}
    \begin{pmatrix}
        -F^{\mu\alpha}F^{\nu}_{\ \alpha}\\
        c_2g^{\mu\nu} F^2\\
        \bar{\psi}i\overleftrightarrow{\mathcal{D}}^{(\mu}\gamma^{\nu)}\psi
    \end{pmatrix}_R
    = \lim_{\epsilon\to0}
    \begin{pmatrix}
        Z_{gg} & (Z_{gg}-Z_F)/4 & Z_{gq} \\
        0 & Z_{F} & 0\\
        Z_{qg} & Z_{qg}/4 & Z_{qq}
    \end{pmatrix}
    \begin{pmatrix}
        -F^{\mu\alpha}F^{\nu}_{\ \alpha}\\
        c_2g^{\mu\nu} F^2\\
        \bar{\psi}i\overleftrightarrow{\mathcal{D}}^{(\mu}\gamma^{\nu)}\psi
    \end{pmatrix}
\end{equation}
Comparing its third row with Eq.~(\ref{eq:Hatta-R}) yields $4Z_B-Z_Q=0$ as an exact identity of perturbative series at every loop order. By repeating the previous calculation of the trace of the renormalized quark EMT, one obtains,
\begin{align}
    g_{\mu\nu}T_{q,R}^{\mu\nu} 
    &= \lim_{\epsilon\to0} \left[-Z_Q+d c_2 Z_B\right]F^2 \\
    &= \lim_{\epsilon\to0} \left[-Z_Q+4Z_B\right]F^2 =0 .
\end{align}
This proves that $x=0$ for $c_2=4/d$ to all orders in perturbation theory. And the trace of the gluon EMT then contains the full trace anomaly $\beta(g)/(2g)$.

\section{Symmetric Rank-4 Tensor} \label{appsec:rank4}

Consider the parity-even, charge-conjugation-even, gauge-invariant dimension-six operators in pure Yang–Mills theory with four fully symmetric free indices, with ${\cal D}_\mu = \partial_\mu - igA_\mu$. We retain equation-of-motion operators. Total-derivative and gauge-variant (BRST-exact) operators can appear as counterterms but do not mix back into this sector, so we omit them. A symmetric rank-four tensor decomposes into a symmetric traceless rank-four tensor, a metric times a symmetric traceless rank-two tensor, and a double metric times a scalar. Let $\mathrm{ST}_r$ denote the symmetric traceless projection of rank $r$, defined in $d$ dimensions before renormalization. 

The available field structures are ${\cal D}^2F^2$ and $F^3$; operators linear in $F$ vanish because the generators are traceless. Integration by parts, the antisymmetry of $F_{\mu\nu}$, the Bianchi identity and $[{\cal D}_\mu, {\cal D}_\nu] = -igF_{\mu\nu}$ reduce each sector to the representatives: 
\begin{align}
    Q_4^{\mu\nu\rho\sigma} &={\rm ST}_4\!\left[F^{\mu\alpha}{\cal D}^\nu {\cal D}^\rho F^\sigma{}_{\alpha}\right], \label{eq:Bspin4}\\
    Q_{2,1}^{\mu\nu} &={\rm ST}_2\!\left[({\cal D}^\mu F^a_{\alpha\beta})({\cal D}^\nu F^{a\alpha\beta})\right],\nonumber\\
    Q_{2,2}^{\mu\nu} &={\rm ST}_2\!\left[({\cal D}_\alpha F^{a\mu\alpha})({\cal D}_\beta F^{a\nu\beta})\right],\nonumber\\
    Q_{2,3}^{\mu\nu} &={\rm ST}_2\!\left[({\cal D}^\nu F^{a\mu\alpha})({\cal D}^\beta F^a_{\alpha\beta})\right], \label{eq:Bspin2}\\
    Q_{0,1} &=g f^{abc} F^a_{\alpha\beta}F^{b\alpha}{}_{\gamma}F^{c\beta\gamma},\nonumber\\
    Q_{0,2} &=(D_\alpha F^{a\alpha\lambda})({\cal D}^\beta F^a_{\beta\lambda}). 
    \label{eq:Bspin0}
\end{align}
In the spin-four sector, each field strength must carry one free index, since a field strength with two free indices is antisymmetric in them; this leaves the single structure in Eq.~(\ref{eq:Bspin4}). In the rank-two sector, the $F^3$ structure is not independent:
\begin{equation}
    \mathrm{ST}_2[g f^{abc}F^{a\,\mu\alpha}F^{b}{}_{\alpha\beta}F^{c\,\beta\nu}] \simeq -\frac{1}{2} Q_{2,1}^{\mu\nu} - Q_{2,3}^{\mu\nu} + (\text{total derivatives})
\end{equation}
which follows from evaluating $F^{a\,\mu\beta}[{\cal D}^\alpha, {\cal D}^\nu]F^a_{\alpha\beta}$ by integration by parts. The operators $Q_{2,2}$, $Q_{2,3}$ and $Q_{0,2}$ vanish by the EOM ${\cal D}_\alpha F^{a\,\alpha\mu}=0$. Embedding the lower-rank operators with metrics gives six rank-four multiplets: 
\begin{equation}
    \left\{Q_4^{\mu\nu\rho\sigma},\;
    g^{(\mu\nu}Q_{2,i}^{\rho\sigma)},\;
    g^{(\mu\nu}g^{\rho\sigma)}Q_{0,j}\right\}.
    \label{eq:Bembedded}
\end{equation}
with $i=1,2,3$ and $j=1,2$. The multiplicities of the inequivalent Lorentz representations are thus $(m_4,m_2,m_0)=(1,3,2)$. Lorentz covariance in $d$ dimensions forbids mixing between inequivalent representations. In the above irreducible basis, the renormalization matrix is therefore block diagonal,
\begin{equation}
 Z_{\rm irr}={\rm diag}\left(Z_4,Z_2^{3\times3},Z_0^{2\times2}\right)
 \label{eq:Bblocks}
\end{equation}
There are at most $1^2+3^2+2^2 = 14$ coefficients. Each block is renormalized independently. 

A reducible basis may instead use operators without trace subtractions, 
\begin{equation}
    \left\{ \tilde Q_4^{\mu\nu\rho\sigma},\ g^{(\mu\nu}\tilde Q^{\rho\sigma)}_{2,i},\ g^{(\mu\nu}g^{\rho\sigma)}Q_{0,j} \right\}
\end{equation}
where $\tilde Q_4$ and $\tilde Q_{2,i}$ are defined as $Q_4$ and $Q_{2,i}$ with the projection $\mathrm{ST}_r$ replaced by symmetrization $\mathrm{S}$ only. The operator $\tilde Q_4$ contains spin-four, spin-two and scalar components, and each $\tilde Q_{2,i}$ contains spin-two and scalar components. Covariance still forbids an operator from mixing into a representation it does not contain, so the renormalization matrix is block triangular,
\begin{equation}
    Z_{\rm red} = 
    \begin{pmatrix} 
    Z_{44} & Z_{42} & Z_{40}\\ 
    0 & Z_{22} & Z_{20}\\ 
    0 & 0 & Z_{00}
    \end{pmatrix}
\end{equation}
with blocks of size $1\times1$, $1\times3$, $1\times2$, $3\times3$, $3\times2$ and $2\times2$. It contains up to $1+3+2+9+6+4=25$ coefficients. The 11 entries in $Z_{42}$, $Z_{40}$ and $Z_{20}$ mix different Lorentz representations and are absent in the irreducible renormalization.

\bibliographystyle{apsrev4-1}
\bibliography{ref}

\end{document}